\documentclass[9pt,twocolumn,twoside]{osajnl}

\journal{josab} 

\setboolean{shortarticle}{false}

\usepackage{bm}
\usepackage{braket}

\title{Birefringent atomic vapor laser lock in a hollow cathode lamp}

\author[1,*]{Takumi Sato}
\author{Yusuke Hayakawa}
\author{Naohiro Okamoto}
\author{Yusuke Shimomura}
\author{Takatoshi Aoki}
\author{Yoshio Torii}

\affil[1]{Institute of Physics, The University of Tokyo, 3-8-1 Komaba, Meguro-ku, Tokyo 153-8902, Japan}

\affil[*]{Corresponding author: t-sato@g.ecc.u-tokyo.ac.jp}
\setboolean{displaycopyright}{true}

\begin{abstract}
We report a robust method of stabilizing a laser to the frequency of an atomic transition using a hollow cathode lamp. In contrast to the standard dichroic atomic vapor laser lock (DAVLL) method, which uses dichroism induced by a longitudinal magnetic field, we employ birefringence induced by a transversal magnetic field. We applied this method to the $(5s^2)\ {}^{1}S_{0} - (5s5p)\ {}^{1}P_{1}$ transition (461 nm) of Sr. Although the hollow cathode is made of ferromagnetic material, we successfully applied a magnetic field of sufficient strength to obtain an error signal with a theoretical maximum slope. This method may be applied to other hollow cathode lamps of different atomic species.
\end{abstract}

\begin{document}
\maketitle

\section{Introduction}
Laser frequency stabilization by spectroscopy of atomic vapors is essential for laser cooling of neutral atoms. Alkali atoms such as Rb and Cs are typical atomic species used for laser cooling experiments because they have a relatively high saturation vapor pressure, enabling laser spectroscopy with a room-temperature vapor cell. Alkaline-earth atoms are other atomic species used for laser cooling experiments. However, spectroscopy with a standard vapor cell is not applicable because they have a relatively low vapor pressure at room temperature (e.g., $\sim 10^{-19}$ Torr for Sr).

Hollow cathode lamps (HCL) are useful for spectroscopy of alkaline-earth atoms because, with moderate power consumption, they provide sufficient vapor pressure in a discharge \cite{Aoki_2012,simple461nmlasersystem,Yb}. Furthermore, one obtains not only neutral atoms in the ground state but also atoms in metastable states \cite{oreno,707} and ions \cite{Ca,davll2,Jung_2017} in the HCL.

Dichroic atomic-vapor laser lock (DAVLL) \cite{davll5,davll,davll3} is known as a simple and robust method for  modulation-free laser frequency stabilization over a wide range in locking frequencies. In this method, the slope of the error signal is optimized when the Zeeman shift induced by the applied longitudinal magnetic field is close to the Doppler width. The applications of the DAVLL method for Yb \cite{davll4}, $\mathrm{Yb}^+$ \cite{davll2} and Ga \cite{Ga} in HCLs have been reported. In these studies, the homogeneity of the magnetic field in the HCL deteriorates because the hollow cathode, which is made of ferromagnetic material, provides magnetic shielding \cite{davll6}. Generally, the magnetic field inside a cylindrical ferromagnetic material in a longitudinal magnetic field is complicated \cite{shielding,shielding2}. Therefore, tuning the strength of the applied magnetic field becomes difficult when optimizing the DAVLL signal in a HCL.

As an alternative method to DAVLL, modulation-free laser frequency stabilization using a transversal magnetic field was reported by Hasegawa and Deguchi \cite{t-davll}, where birefringence, instead of dichroism, was used to obtain an error signal for laser frequency stabilization. We call this method birefringent atomic-vapor laser lock (BAVLL) to emphasize its physical origin. The advantage of BAVLL over DAVLL is that the slope of the error signal is less dependent on the strength of the magnetic field \cite{t-davll}. Moreover, a uniform transversal magnetic field yields an almost uniform magnetic field inside the cylindrical hollow cathode \cite{shielding2}. In Ref. \cite{t-davll}, BAVLL was demonstrated for Ba${}^+$ in a discharge cell, the electrodes of which were made of nonmagnetic titanium alloy. However, a demonstration of BAVLL in a HCL has not been reported.

 In this paper, we report the first demonstration of BAVLL in a HCL with the Sr $(5s^2)\ {}^1S_0-(5s5p)\ {}^1P_1$ (461 nm) transition. We find the applied magnetic field is significantly shielded by the hollow cathode below 0.15 T, above which the applied magnetic field begins to penetrate. When we apply a magnetic field of 0.32 T, the magnetic field inside the hollow cathode reaches 0.1 T and the slope of the BAVLL signal reaches its theoretical maximum. We stabilize the frequency of a 461-nm laser to a BAVLL system using a HCL and find that the frequency variations ($\sim 1$ MHz peak-to-peak over 20 h) are well below the natural width of the 461-nm transition (30 MHz). This method leads to a significant simplification of laser cooling experiments using Sr and other alkaline-earth atoms.

\section{Experimental Setup}

\begin{figure}[t]
	\begin{center}
		\includegraphics[width=95mm]{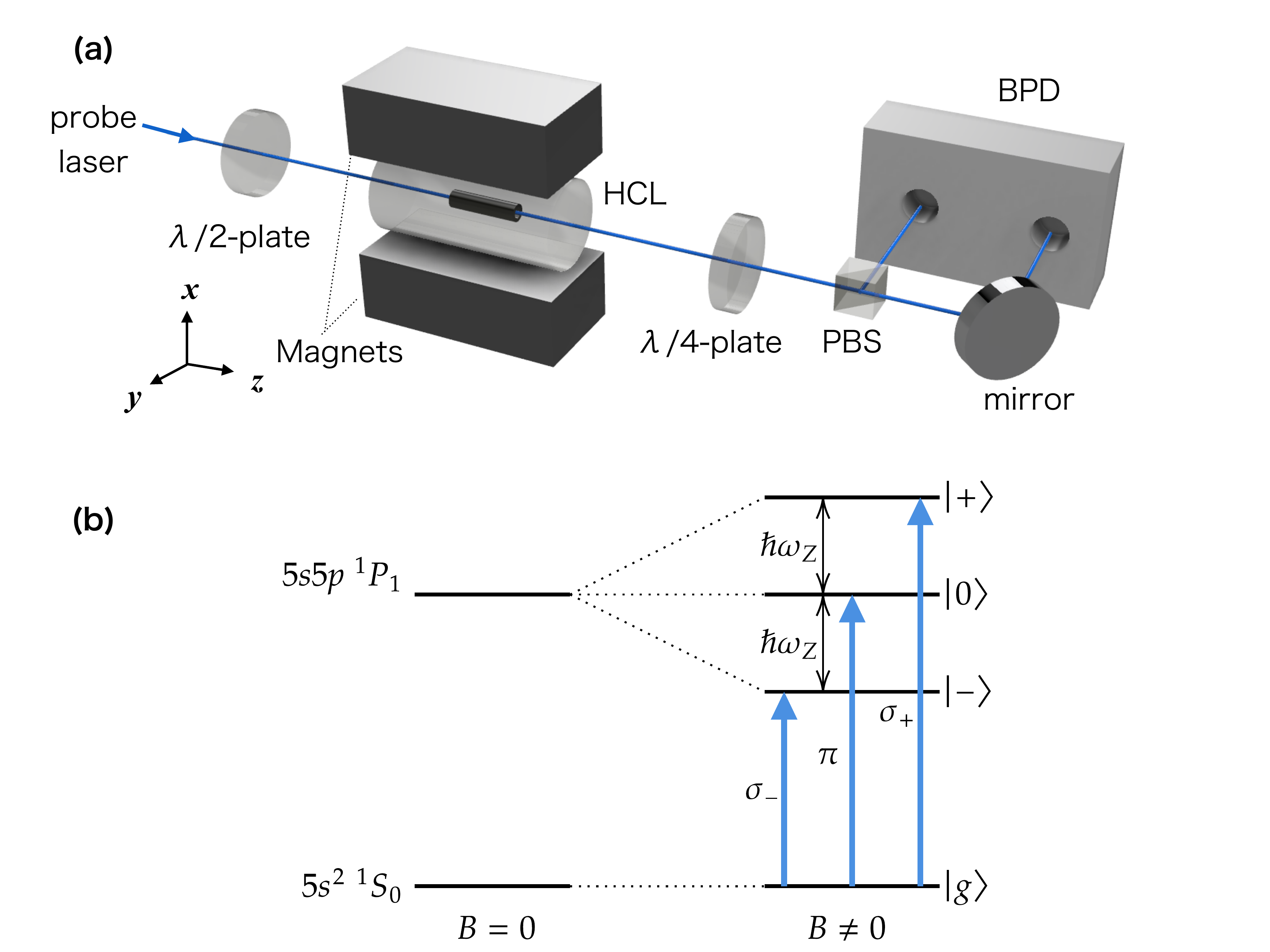}
		\caption{(a)Schematic diagram of the experimental setup. HCL, hollow cathode lamp; PBS, polarizing beam splitter; BPD, balanced photodiode. (b)Relevant energy levels of Sr. $\ket{g}$, $\ket{0}$ and $\ket{\pm}$ represent ($5s^2$) $^1S_0$, ($5s5p$) $^1P_1$,  $M_J=0$ and ($5s5p$) $^1P_1$, $M_J=\pm1$ states, respectively.}
		\label{fig:expsetup}
	\end{center}
\end{figure}

In the experimental setup [Fig. \ref{fig:expsetup}(a)], a see-through Sr HCL (Hamamatsu L2783-38NE-SR), filled with a 3.5 Torr Ne buffer gas, is placed in a vertical magnetic field provided by a pair of two neodymium magnets, the residual magnetic flux density of which is 1.2 T and the dimensions in the $x, y, z$-directions are 20 mm, 30 mm, 60 mm, respectively. The strength of the magnetic field may be varied between 0.14 T to 0.32 T by adjusting the distance of the two magnets. The hollow cathode, which is made of ferromagnetic material, has a length of 20 mm and a bore diameter of 3 mm. The discharge current is fixed at 6 mA, which offers a maximum absorption of 30\% for the 461 nm transition.

A 461 nm probe laser beam (140 $\mu$W, 1 mm $1/e^2$-radius) is derived from a blue external-cavity laser diode \cite{simple461nmlasersystem} and sent through the hollow cathode. The polarization of the probe beam is tilted by $45^\circ$ with respect to the direction of the applied magnetic field with a $\lambda/2$-plate before entering the HCL such that the $x$ and $y$ components of the probe beam have the same intensity. If we define the quantization axis in the direction of the magnetic field ($x$ axis), the $\pi$ transition ($\ket{g} \leftrightarrow \ket{0}$) occurs for the $x$ component and the $\sigma_+$ and $\sigma_-$ transitions ($\ket{g} \leftrightarrow \ket{\pm}$) occur for the $y$ component [see Fig. \ref{fig:expsetup}(b)].

After passing through the HCL, the polarization of the probe beam is elliptical because the $x$ and $y$ components of the probe beams have different resonance frequencies because of the Zeeman shift, and the Sr vapor in the hollow cathode exhibits birefringence. The ellipticity of the probe is then analyzed by a $\lambda/4$-plate, a polarizing beam splitter, and a balanced photodetector (BPD) to obtain a BAVLL signal (see Appendix for details).


\section{Results and Discussion}

\begin{figure}[hhhh]
	\begin{center}
		\includegraphics[width=85mm]{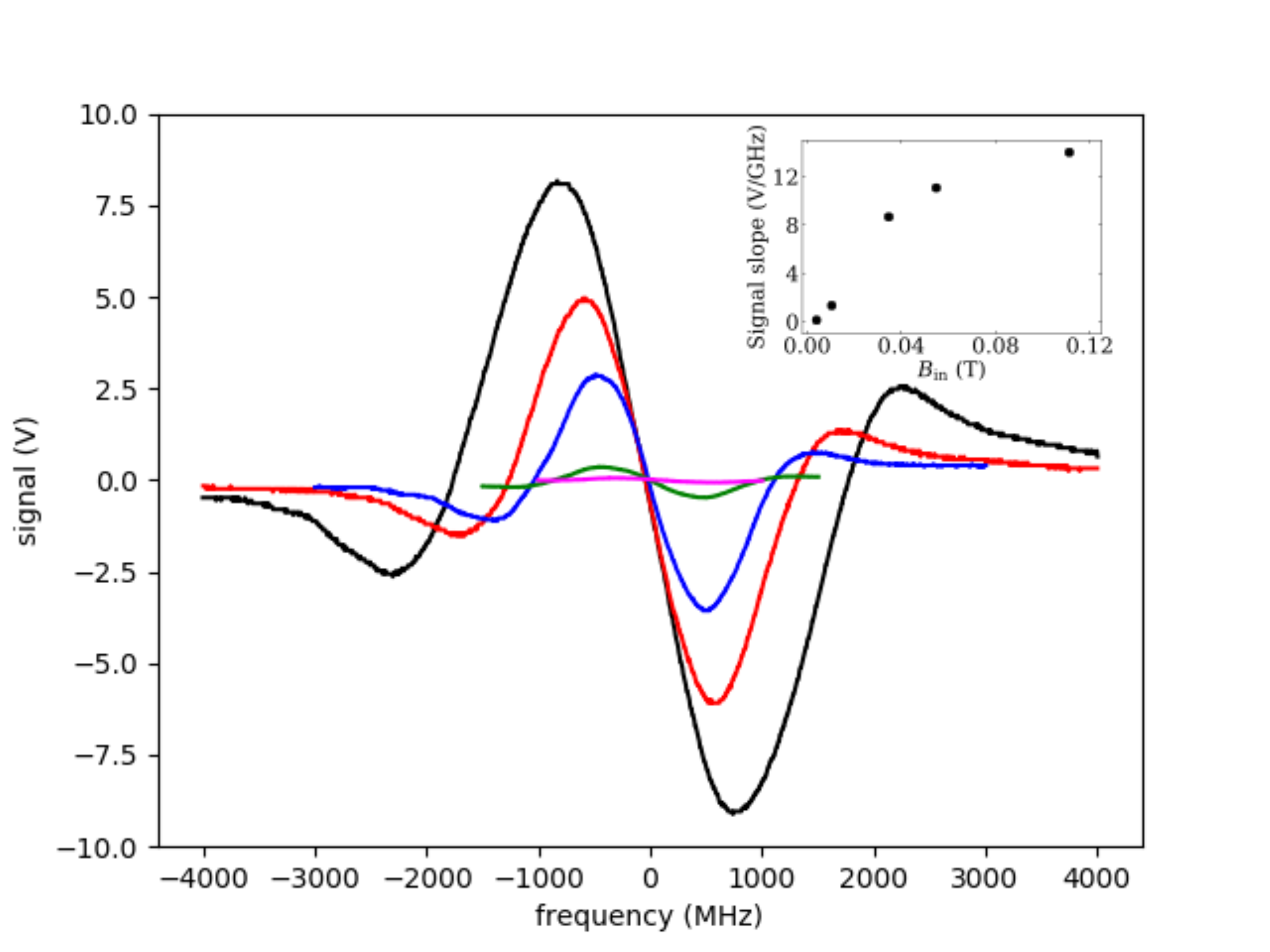}
		\caption{BAVLL signals obtained in experiments with applied magnetic fields of 0.14 T, 0.18 T, 0.24 T, 0.28 T, and 0.32 T. The signal amplitude increases with increasing the applied magnetic field. The inset shows the dependence of the signal slope on the magnetic field strength $B_\mathrm{in}$ inside the hollow cathode (see text and Fig. \ref{fig:result3}).}
		\label{fig:result4}
	\end{center}
\end{figure}

\begin{figure}[tttt]
	\begin{center}
		\includegraphics[width=75mm]{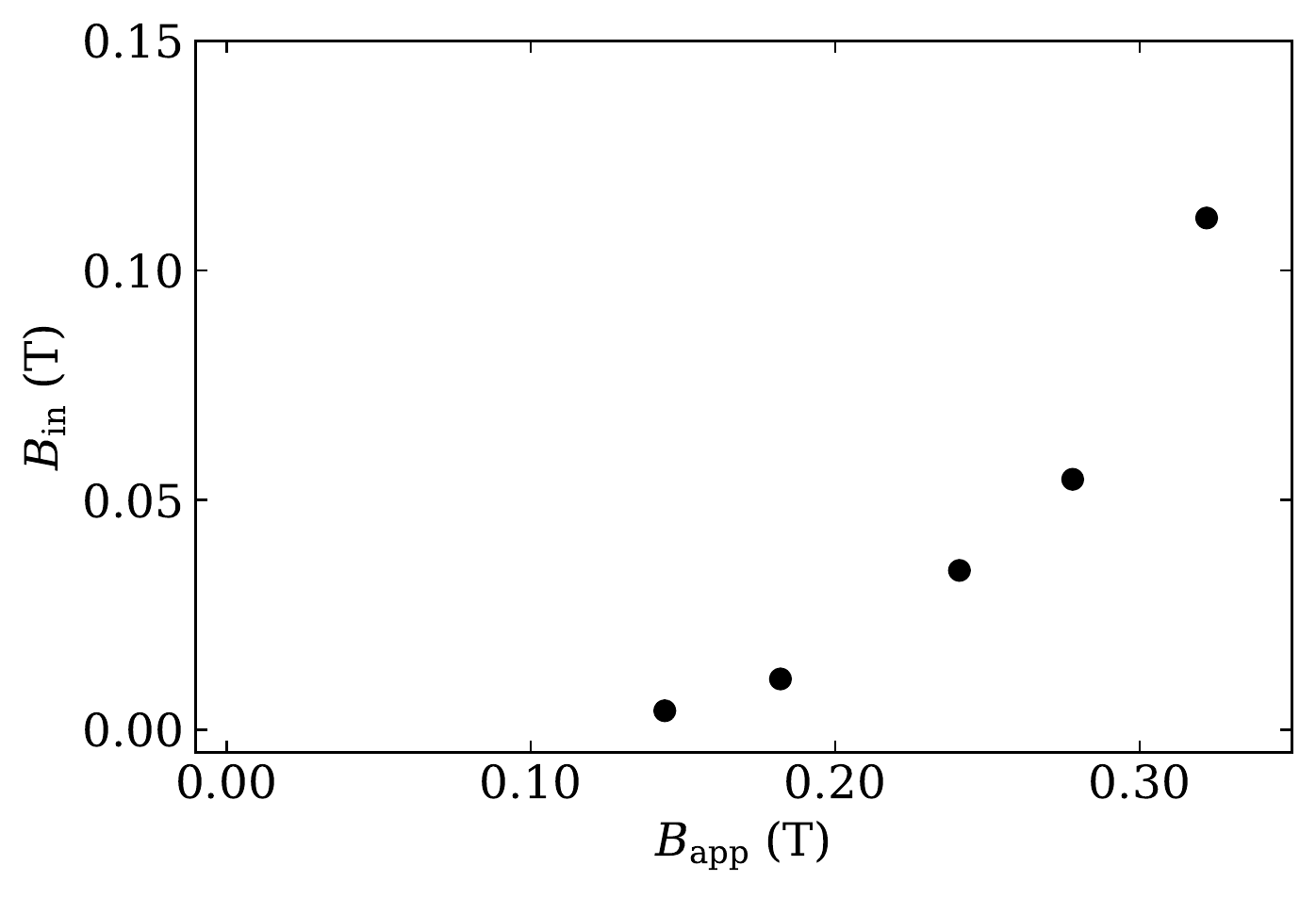}
		\caption{Dependence of the magnetic field strength $B_\mathrm{in}$ inside the hollow cathode on the applied magnetic field strength $B_\mathrm{app}$.}
		\label{fig:result3}
	\end{center}
\end{figure}

As expected from the theoretical curve [Eq. (\ref{eq:1})], the BAVLL signals obtained for applied magnetic fields varying from 0.14 T to 0.32 T (Fig. \ref{fig:result4}) show a Doppler-broadened dispersive profile reflecting Zeeman-induced birefringence. By fitting the theoretical curve, we deduce for each signal the Zeeman shift $\omega_Z=\mu_\mathrm{B} B/\hbar$, from which we can infer the magnetic fields inside the hollow cathode.

Figure \ref{fig:result3} shows the dependence on the applied magnetic field $B_{\mathrm{app}}$ of the magnetic field $B_{\mathrm{in}}$ inside the hollow cathode. The applied magnetic field is significantly shielded below 0.15 T, above which it begins to penetrate into the hollow cathode. The dependence of the slope of the BAVLL signal on the magnetic field strength inside the hollow cathode is shown in the inset of Fig. \ref{fig:result4}. The slope is less dependent on $B_{\mathrm{in}}$ above 0.1 T.

As shown in Fig. \ref{fig:slope} in Appendix, the slope of the theoretical BAVLL signal reaches its maximum at $\omega_Z \sim 1.5 \omega_D$ ($\omega_D=2\pi \times 0.7$ GHz is the Doppler width) and remains almost constant when $\omega_Z > 1.5 \omega_D$. Therefore, if the strength of the magnetic field inside the hollow cathode is $> 1.5\hbar\omega_D/\mu_B=0.07\ \mathrm{T}$, the slope remains almost constant. This estimation is consistent with the experimental results shown in the inset of Fig. \ref{fig:result4}.

We employed a BAVLL system with an applied magnetic field of 0.32 T for laser frequency stabilization. To monitor the frequency stability of the laser lock, we stabilized two separate 461 nm lasers: one was locked to a BAVLL system, the other to a system performing Doppler-free polarization spectroscopy \cite{simple461nmlasersystem}.  Because the width of the error signal of the former ($\sim$ 1.5 GHz, see Fig. \ref{fig:result4}) is more than ten times wider than that of the latter  ($\sim$ 100 MHz, see \cite{simple461nmlasersystem}), the latter is effectively regarded as an absolute frequency reference. The error signal of each system was fed back both to the PZT and the diode current. For the BAVLL system, we used a Wollaston prism as a polarizing beam splitter in front of the balanced photodetector. Part of the light from each laser was combined at a beam splitter and detected by a photodetector to obtain the beat signal. The frequency of the beat signal, which corresponds to the frequency difference of the two lasers, was measured by a frequency counter. We rotated the $\lambda/4$-plate of the BAVLL system such that the beat frequency was set to approximately 25 MHz.

\begin{figure}[t]
	\begin{center}
		\includegraphics[width=85mm]{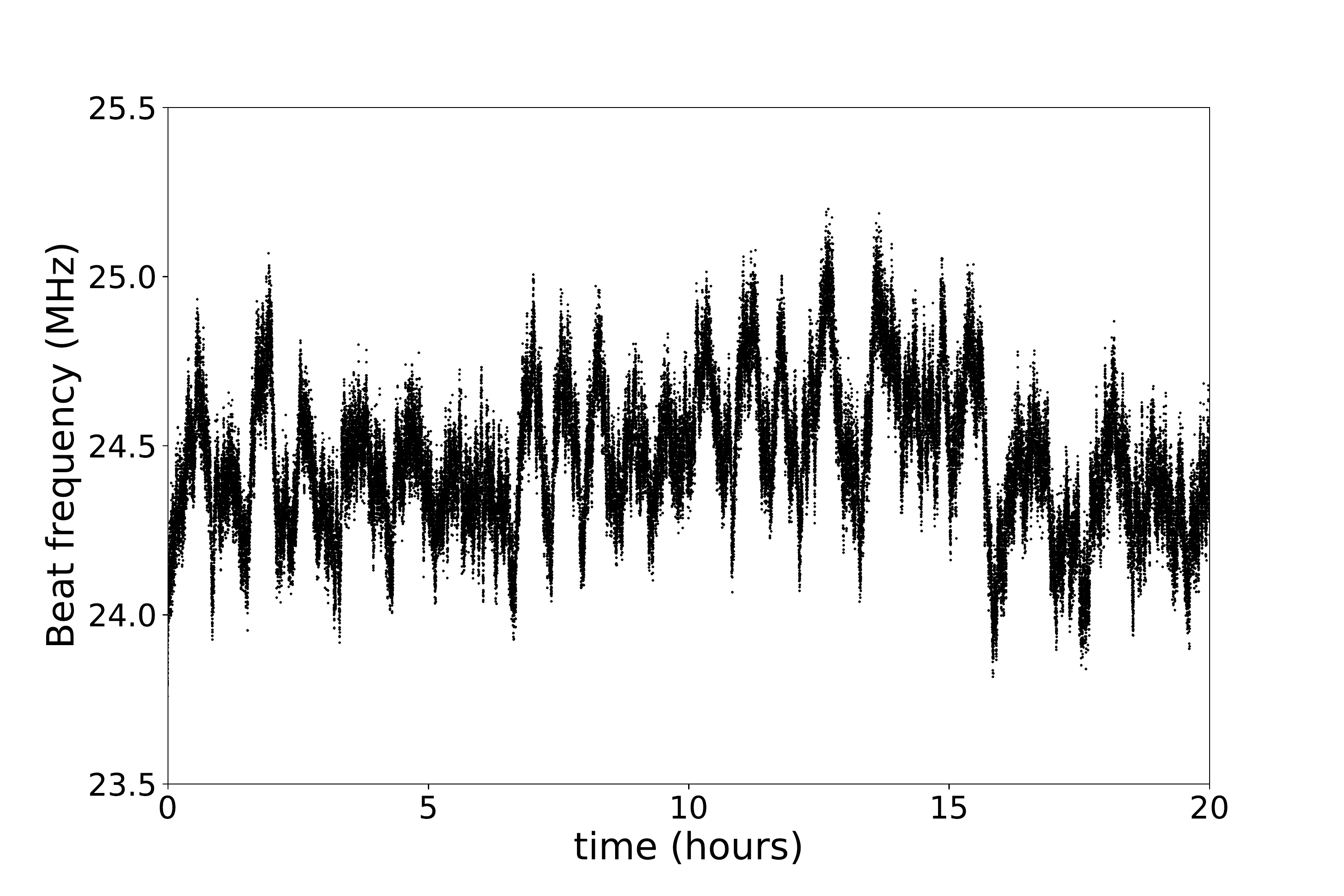}
		\caption{Measured beat frequency between two lasers, each locked to a BAVLL system and a Doppler-free polarization spectroscopy system. The gate time of the frequency counter was 1 s.} 
		\label{fig:result5}
	\end{center}
\end{figure}

Figure \ref{fig:result5} shows the variations in beat frequency over 20 hours. The period of the variations in beat frequency was around 1 h, which was the typical period of the temperature variations ($0.4  \ {}^\circ \mathrm{C}$ peak-to-peak) in the laboratory.  From the Allan deviation of the data in Fig. \ref{fig:result5}, we estimate the amplitude of the beat frequency variations to be 0.2 MHz (0.4 MHz peak-to-peak).  If we assume the stability of the BAVLL lock is limited by the temperature-dependent birefringence of the optical elements, the estimated temperature coefficient of the frequency shift was $ 1.0 \ \mathrm{MHz}/{}^\circ \mathrm{C}$, which is consistent with the result obtained with the DAVLL system \cite{davll}.
 This stability of the laser lock is satisfactory considering the natural width of the 461 nm transition (30 MHz). Indeed, we have employed a 461-nm laser locked to a BAVLL system for Sr magneto-optical trap (MOT) for more than two years.

We note that the BAVLL method may be applied to Sr atoms not only in the ground state but also in the metastable states. We also note that the BAVLL method may be applied to HCLs of other species. Furthermore, because  the slope of the BAVLL signal is less dependent on the strength of the applied magnetic field, the BAVLL method offers a more simple way of laser locking for standard vapor cells of alkali atoms such as Rb and Cs, to which the DAVLL method has mainly been applied.

\section{conclusion}
We demonstrated the BAVLL method of stabilizing a laser frequency to an atomic transition using a Sr hollow cathode lamp in a transversal magnetic field. Despite the shielding effect of the ferromagnetic hollow cathode, we may apply a magnetic field of sufficient strength to obtain an error signal with a theoretical maximum slope. A 461-nm laser locked to a BAVLL system showed frequency variations of $\sim 1$ MHz over a 20-h period, which is well below the natural width of the transition. This simple and robust method may be extended to other HCLs of atoms with low vapor pressures and standard vapor cells of alkali atoms and contributes in simplifying the laser cooling of these atoms.

\section*{Appendix}
\appendix
\section*{Birefringent Atomic Vapor Laser Lock Signal}
The theory of the BAVLL signal was fully explored in \cite{t-davll}. Here, we summarize the essence of the theory for readers' convenience.

We consider the interaction of an atomic vapor with a linearly polarized laser beam propagating in the $z$-direction [Fig. \ref{fig:expsetup}(a)]. We assume that the total angular momentum quantum number of the ground (excited) state is $J=0$ ($J=1$) and the $M_J=\pm 1$ states of the excited state are Zeeman shifted by $\pm \hbar \omega_{Z}$ ($\hbar$ the reduced Planck constant) [see Fig. \ref{fig:expsetup}(b)]. For Sr, the $g$-factor of the ${}^1P_1$ state is one and the Zeeman shift is given by $\hbar\omega_Z=\mu_\mathrm{B} B$, where $\mu_\mathrm{B}$ is the Bohr magneton and $B$ the magnetic field strength. The polarization of the probe beam is tilted by $45^\circ$ with respect to the $x$-axis. A magnetic field is applied in the direction of $x$-axis, which we choose as the quantization axis. The $x$ component of the probe beam induces the $\pi$ transition while the $y$ component induces both the $\sigma_+$ and $\sigma_-$ transitions. If the Doppler width is much wider than the natural line width, the real and imaginary parts of the electric susceptibilities $\chi_{x,y} \equiv \chi'_{x,y}+i\chi''_{x,y}$ are described as \cite{t-davll}

\begin{eqnarray}
\chi'_x \!\!\!\!&=&\!\!\!\!-aF\left(\frac{\delta}{\omega_D}\right),\\
\chi'_y \!\!\!\!&=&\!\!\!\!-\frac{1}{2}a\left[F\left(\frac{\delta-\omega_Z}{\omega_D}\right)+F\left(\frac{\delta+\omega_Z}{\omega_D}\right)\right],\\
\chi''_x \!\!\!\!&=&\!\!\!\!aG\left(\frac{\delta}{\omega_D}\right),\\
\chi''_y \!\!\!\!&=&\!\!\!\!\frac{1}{2}a\left[G\left(\frac{\delta-\omega_Z}{\omega_D}\right)+G\left(\frac{\delta+\omega_Z}{\omega_D}\right)\right],
\end{eqnarray}
where $G(\xi)=\exp(-\xi^2)$ represents the Gauss function, $F(\xi)=(2/\sqrt{\pi})\exp(-\xi^2)\int_{0}^{\xi}\exp(\eta^2)d\eta$ the Hilbert transform of $G(\xi)$, $\omega_D=k\sqrt{2k_\mathrm{B}T/M}$ the Doppler width with $k$ the wave number, $k_\mathrm{B}$ the Boltzmann constant, $T$ the temperature, and $M$ the mass of the atom, and $\delta$ the detuning from the resonance in the absence of the magnetic field. We define $\chi''_{\mathrm{max}}=aG(0)=a$ as the maximum of $\chi''_{x}$, which is determined in an experimental analysis from the relation
\begin{eqnarray}
kL\chi''_{\mathrm{max}} &=& \mathrm{OD}_{\mathrm{max}},
\end{eqnarray}
\begin{figure*}[hhhh]
	\begin{center}
		\includegraphics[width=\linewidth]{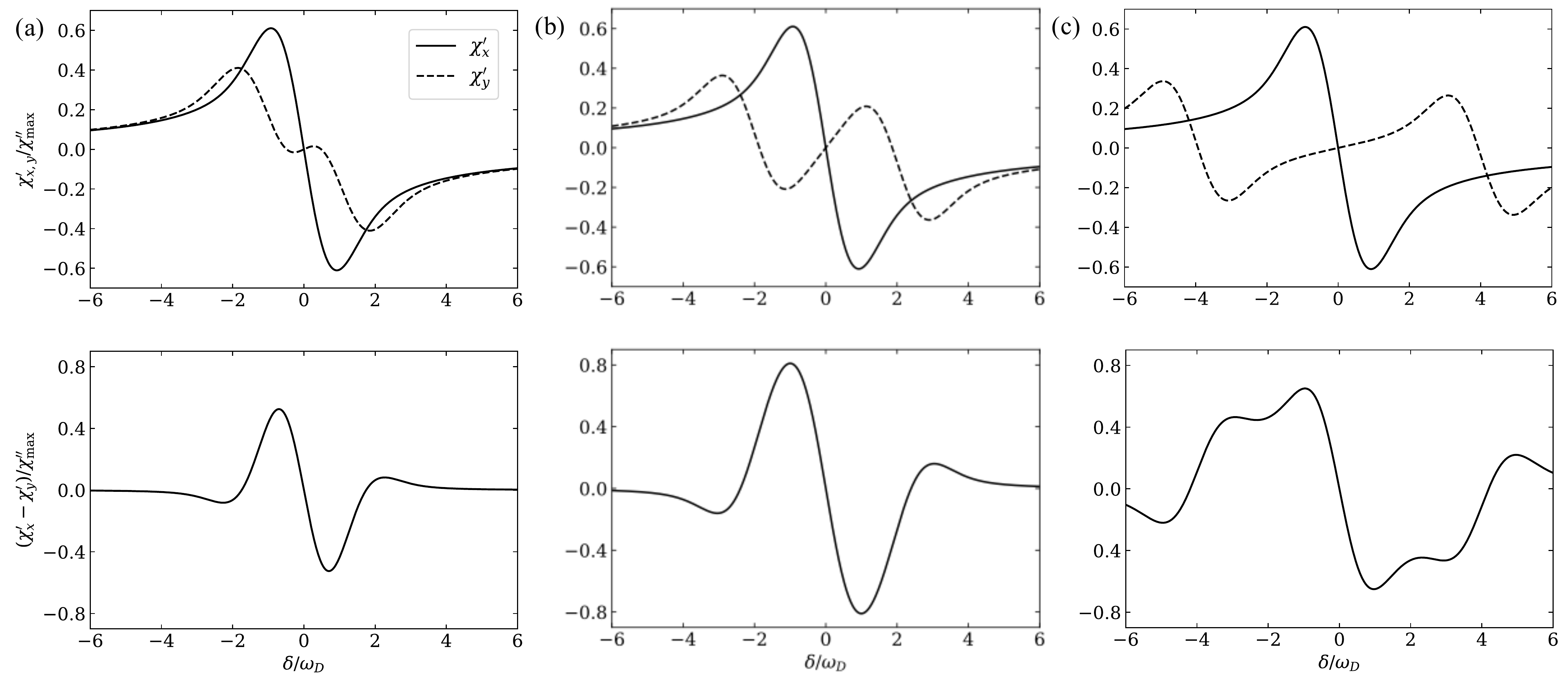}
		\caption{Real parts of the susceptibilities $\chi_{x,y}'$ (upper traces) and difference between $\chi'_x$ and $\chi'_y$ (lower traces) when (a) $\omega_Z/\omega_D=1$, (b) $\omega_Z/\omega_D=2$, and (c) $\omega_Z/\omega_D=4$.}
		\label{fig:chi}
	\end{center}
\end{figure*}
where $L$ denotes the interaction length of the atomic vapor and $\mathrm{OD_{\mathrm{max}}}$ the maximum optical density of the observed Doppler broadened absorption profile without the magnetic field. The upper traces of Fig. \ref{fig:chi} show the shape of $\chi_{x,y}'$ for $\omega_Z/\omega_D = 1, 2,$ and $4$.

Using the Jones calculus, the polarization of the input probe beam with amplitude $E_0$ is expressed as
\begin{eqnarray}
\bm{E}_{\mathrm{in}}=\frac{E_0}{\sqrt{2}}
\begin{pmatrix} 1\\ 1 \end{pmatrix}.
\end{eqnarray}
In the limit of weak absorption, the polarization of the probe beam after passing through the atomic vapor is described as
\begin{eqnarray}
\bm{E}_{\mathrm{out}}&=&\frac{E_0}{\sqrt{2}}
\begin{pmatrix} e^{i\phi}\\ 1 \end{pmatrix},\\
\phi&=&\frac{kL}{2}\left(\chi'_x-\chi'_y\right),
\end{eqnarray}
where $\phi$ is the phase difference between the $x$ and $y$ components. The output beam is then passed through a quarter-wave plate, the fast axis of which is rotated by $45^\circ$ with respect to the $x$ axis, resulting in 
\begin{eqnarray}
\begin{pmatrix} E_x \\ E_y \end{pmatrix}
&=&\frac{1}{\sqrt{2}}
\begin{pmatrix} e^{-i\pi/4} &e^{i\pi/4} \\ e^{i\pi/4}&e^{-i\pi/4}
\end{pmatrix}
\bm{E}_\mathrm{out} \nonumber \\
&\simeq&\frac{E_0}{\sqrt{2}}\exp\left( {\frac{i\phi}{2}}\right)
\begin{pmatrix} 1+\phi/2 \\ 1-\phi/2 \end{pmatrix},
\end{eqnarray}
where we assume $\phi\ll 1$.

To obtain the BAVLL signal, we observe the intensity difference between the $x$ and $y$ components, which is
\begin{eqnarray}
I_x-I_y&=&\frac{1}{2}\epsilon_0c(|E_x|^2-|E_y|^2) \nonumber \\
&=&\phi I_0 \nonumber \\
&=&\frac{kL}{2}(\chi_x'-\chi_y')I_0,
\label{eq:1}
\end{eqnarray}
where $I_0=\epsilon_0 cE_0^2/2$ is the total intensity of the probe beam. The lower traces of Fig. \ref{fig:chi} show $\chi'_x-\chi'_y$ for $\omega_Z/\omega_D = 1, 2,$ and $4$. 

With a maximum of $\chi'_x$ of about $0.6\chi''_{\mathrm{max}}$, the maximum intensity difference $I_{\mathrm{max}}$ of the BAVLL signal in the limit of $\omega_Z \gg \omega_D$, where $\chi'_y \sim 0$ around $\delta/\omega_D=0$, is therefore given by
\begin{eqnarray}
I_{\mathrm{max}}&\simeq&\frac{kL}{2} \times 0.6 \chi''_\mathrm{max}I_0 \nonumber \\
&=& 0.3\times \mathrm{OD}_{\mathrm{max}}I_0.
\end{eqnarray}
If $\mathrm{OD}_{\mathrm{max}}=0.1$, for example, the amplitude of the central dispersive signal corresponds to 3 \% of the total probe intensity. 

\begin{figure}[t]
	\begin{center}
		\includegraphics[width=75mm]{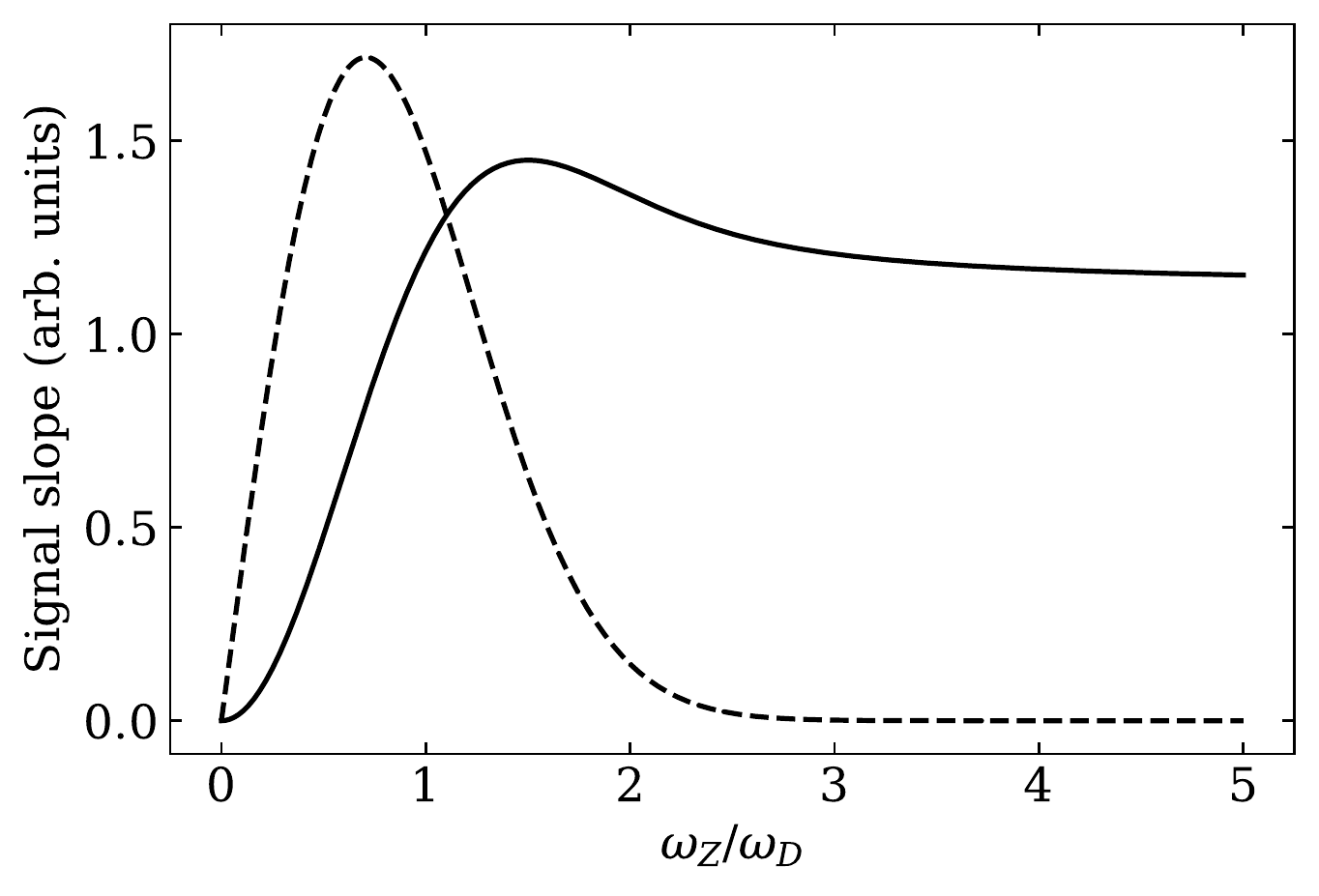}
		\caption{Dependence of the signal slope on $\omega_Z / \omega_D$ at $\delta=0$ for both BAVLL (solid line) and DAVLL (dashed line).}
		\label{fig:slope}
	\end{center}
\end{figure}

The feature of the error signal that is relevant for laser frequency stabilization is not the signal amplitude but the slope of the signal. Figure \ref{fig:slope} shows the signal slope at $\delta=0$ as a function of $\omega_Z / \omega_D$ for both the BAVLL and the DAVLL signals, which is essentially the same as Fig. 4 in \cite{t-davll}. The slope of the DAVLL signal has its maximum at $\omega_Z=0.7\omega_D$, which requires fine tuning of the applied magnetic field. In contrast, the slope of the BAVLL signal has its maximum at $\omega_Z / \omega_D \sim 1.5$, above which the slope stays almost constant. This feature eliminates the need for fine tuning of the applied magnetic field. Moreover, the uniformity of the applied magnetic field is less critical for BAVLL. Therefore, BAVLL is superior to DAVLL, especially when we probe the vapor in a ferromagnetic hollow cathode.

\begin{backmatter}


\bmsection{Funding}
KAKENHI, Japanese Society for the Promotion of Science (JSPS) JP15H02027, JP21H01089.
\bmsection{Disclosures}
The authors declare no conflicts of interest.
\bmsection{Data Availability}
Data underlying the results presented in this paper are not publicly available at this time but may be obtained from the authors upon reasonable request.

\end{backmatter}

\bibliography{bib.bib}


\end{document}